\documentclass[acmtog]{acmart}
\AtBeginDocument{%
  }

\setcopyright{acmlicensed}
\copyrightyear{2025}
\acmYear{2025}
\acmDOI{XXXXXXX.XXXXXXX}

\usepackage{fvextra}
\DefineVerbatimEnvironment{Code}{Verbatim}{breaklines=true}
\usepackage{geometry}
\usepackage{multicol}
\usepackage{subfigure}
\usepackage{subcaption}
\usepackage{longtable}

\begin{document}

\title{THE HIDDEN AI RACE: TRACKING ENVIRONMENTAL COSTS OF INNOVATION}


\author{Shyam Agarwal}
\email{shyagarwal@ucdavis.edu}
\affiliation{%
  \institution{Department of Computer Science, University of California, Davis}
  \country{USA}
}

\author{Mahasweta Chakraborti}
\email{mchakraborti@ucdavis.edu}
\affiliation{%
  \institution{Department of Communication, University of California, Davis}
  \country{USA}
}








\renewcommand{\shortauthors}{Agarwal et al.}

\begin{abstract}
The past decade has seen a massive rise in the popularity of Artificial Intelligence systems, mainly owing to the developments in the generative AI space, which has revolutionized numerous industries and applications. However, this progress comes at a considerable cost to the environment as training and deploying these models consume significant computational resources and energy and are responsible for large carbon footprints in the atmosphere. In this paper, we study the amount of carbon dioxide released by models across different domains, including Natural Language Processing, Computer Vision, and Audio, over varying time periods. By examining parameters such as model size, repository activity (e.g., commits and repository age), task type, and organizational affiliation, we identify key factors influencing the environmental impact of AI development. Our findings reveal that model size and versioning frequency are strongly correlated with higher emissions, while domain-specific trends show that NLP models tend to have lower carbon footprints compared to audio-based systems. Organizational context also plays a significant role, with university-driven projects exhibiting the highest emissions, followed by non-profits and companies, while community-driven projects show a reduction in emissions. These results highlight the critical need for green AI practices, including the adoption of energy-efficient architectures, optimizing development workflows, and leveraging renewable energy sources. We also discuss a few practices that can lead to a more sustainable future with AI, and we end this paper with some future research directions that could be motivated by our work. This work not only provides actionable insights to mitigate the environmental impact of AI but also poses new research questions for the community to explore. By emphasizing the interplay between sustainability and innovation, our study aims to guide future efforts toward building a more ecologically responsible AI ecosystem.
\end{abstract}




\maketitle

\section{Introduction}
The past decade has been marked by significant developments in AI, especially with the developments in Generative AI. There used to be a time when only a small number of research labs were involved in training neural networks. Fast-forward to 2024, and most industries have already integrated AI capabilities, the number of researchers working in the AI space is rapidly increasing, and students are learning and building AI projects as early as high school.
  
This development has been possible because of the increased accessibility of powerful GPUs, which can now be accessed from the comfort of one's home for very low costs. With such ease of use, Commercial AI products based on generative, multi-purpose AI systems that encourage a unified approach to building ML models have also seen a surge. However, this goal of "generality" comes with its own set of dangers to the environment because training on these GPUs takes a lot of computational resources and energy and releases a large amount of carbon dioxide into the atmosphere.

Recent studies highlight that LLM serving, for example, has begun to exceed training in energy consumption, with operational carbon emissions (emissions due to the energy consumption of these applications, which is usually quantifed as the equivalent carbon dioxide emitted) significantly contributing to the environmental impact \cite{Chien2023ReducingCarbon}. For example, serving a single prompt in ChatGPT generates over 4 grams of $CO_2eq$ \cite{Wong2023ChatGPTCarbonFootprint}, which is over 20 times the emissions of a web search query \cite{Griffiths2020InternetCarbonImpact}. As most developments of AI models aim to beat the previous ones on some chosen metric, it often requires training on more GPUs for a longer time, and that too on larger datasets - all of which means more carbon emissions into the environment. It is important to note that although the carbon emission impact of traditional softwares is well-established \cite{Verdecchia2021GreenIT, Mancebo2021Feetings}, assessing the carbon impact of modern day models, like LLMs, is quite complex, especially because they are trained using High-Performance Computing (HPC) clusters where factors like parallel jobs and shared resources make precise quantification challenging. Even well-established LLM leaderboard benchmarks \cite{Trustbit2024LLMBenchmarks, Arena2024LMLeaderboard, Oobabooga2024Benchmark} focus on metrics like accuracy to gauge model performance on tasks, instead of the carbon footprint of the model.

The situation raises the question of whether our current AI practices are sustainable and highlights the need to scrutinize the factors contributing to such high carbon emissions. In this work, we highlight some interesting trends in carbon emissions based on factors like model size, model popularity, and model repository age among others. We also discuss practices and policies that researchers and ML engineers should adopt to promote a more sustainable future with AI development.

\section{Related Works}
The study of how AI model training and deployment impacts the environment is a relatively under-explored area in the literature. However, this line of study has received a lot of attention in the last few years thanks to the pioneering work by Strubell and others, who provided one of the first comprehensive analyses of environmental costs of training the then large natural language processing models \cite{Strubell2019EnergyAP}. They quantified the energy consumption and carbon emissions associated with training these models, highlighting that training a single model can emit as much carbon as five cars in their lifetimes.

Building on this work, Patterson and others \cite{Patterson2021CarbonEmissions, Patterson2022CarbonFootprint} undertook comparative studies of different models and analyzed the factors contributing to the emissions. They also analyzed the trends in power consumption, revealing that since 2012, the computational power required to train some of the most significant AI models has been doubling every 3.4 months. Several studies have also focused primarily on specific model architectures, breaking down the environmental impact of a model across the different steps in its life cycle. Studies like BLOOM \cite{Luccioni2022EstimatingCF} and Nour \cite{Lakim2022HolisticAssessment} have been phenomenal in highlighting which parts of a model's life cycle contribute the most to the total final carbon emissions.

Since most ML models are deployed in cloud environments, some researchers have focused on cloud-specific emission reduction techniques. \cite{Chien2023ReducingCarbon} have proposed techniques like delayed scheduling and workload shifting to reduce the carbon footprint of cloud-based ML operations and have projected potential carbon savings from these techniques in current and future scenarios. \cite{Dodge2022MeasuringCarbon} have highlighted the importance of accounting for geographical variations in power grid carbon intensity and have proposed methods for reducing carbon emissions by strategically choosing cloud instances and scheduling workloads. \cite{Hanafy2023CarbonScaler} introduced CarbonScaler, an approach that reduces carbon emissions of batch workloads in cloud platforms without extending completion times.

Similarly, to minimize operational carbon emissions, Carbon Explorer offers a comprehensive framework aimed at reducing carbon output in data centers \cite{Acun2023CarbonExplorer}. Ecovisor implements a carbon-efficient virtual energy management system to optimize energy usage \cite{Souza2023Ecovisor}. GAIA focuses on optimizing the cost-efficiency of cloud-based carbon reduction strategies \cite{Hanafy2024GoingGreen}, while Caribou targets carbon emission reductions in serverless applications through geospatial workload shifting \cite{Gsteiger2024Caribou}.

Recent works have also explored more targeted approaches to reducing emissions, such as optimization techniques for inference and specific tasks. Rubei and others \cite{rubei2025promptengineeringimplicationsenergy} investigated how prompt engineering techniques (PETs) can impact the carbon emissions of LLaMA 3 models during code generation tasks. By experimenting with the CodeXGLUE benchmark, they demonstrated that using specific tags to distinguish different parts of prompts can reduce energy consumption without compromising performance. Their findings suggest that prompt engineering could play a significant role in reducing the environmental impact of LLMs during inference.

Nguyen and others \cite{nguyen2024sustainablelargelanguagemodel} extended this line of work by addressing both operational and embodied emissions (emissions during the manufacturing process of the hardware) of LLMs to develop sustainable LLM serving strategies. They characterized the performance and energy consumption of LLaMA models with varying parameter sizes using different types of Nvidia GPUs. By modeling carbon emissions based on energy consumption, grid region carbon intensities, chip area, and memory size, they emphasized the importance of considering both operational and embodied emissions when optimizing sustainable serving systems for LLMs.

The development and serving of LLMs also necessitate powerful hardware resources like NVIDIA HGX \cite{Nvidia2024HGX} and Google TPU \cite{GoogleCloudTPU}, which are built with advanced feature sizes and high-capacity memory to support efficient LLM execution. However, the contribution of such latest high-performing devices to embodied carbon emissions is quite high \cite{Gupta2022ACT, Li2023SustainableHPC, Zhang2023EmbodiedCarbon}. Efforts to address these challenges have explored disaggregated GPU usage, as seen in frameworks like SplitWise, which executes prefill and decode phases separately on different GPUs \cite{Patel2024Splitwise}. Tools like PowerInfer aim to adapt LLMs to consumer-grade GPUs by reducing the LLMs’ need for GPU onboard memory, and hence, reduce environmental impact \cite{Song2023PowerInfer}. Yet, understanding carbon emissions at granular levels—such as per-token emissions across platforms—remains underexplored and critical for sustainable AI development.

Shi and others \cite{shi2024greenllmdisaggregatinglargelanguage} tackled the challenge by particularly focusing on the issue of electronic waste driven by the demand for high-performing GPUs. They proposed GreenLLM, an SLO-aware LLM serving framework that minimizes carbon emissions by reusing older GPUs. GreenLLM disaggregates specific computations onto older GPUs, enabling significant carbon savings while meeting service-level objectives (SLOs). Their evaluations demonstrated up to a 40.6\% reduction in carbon emissions compared to standard LLM serving on newer GPUs, with over 90\% of requests meeting latency requirements across diverse scenarios.

Despite the work done in analyzing the carbon footprints of AI models and devising strategies to reduce them, there needs to be standardized methods and practices to quantify and report these emissions. Several tools have been implemented towards this end. Faiz and others proposed LLMCarbon, an end-to-end carbon footprint projection model for both dense and mixture-of-experts (MoE) large language models that provide a comprehensive approach to estimating carbon emissions during training, inference, experimentation, and storage stages \cite{Faiz2023LLMCarbon}. Lacoste and others introduced an online machine-learning emissions calculator that takes as input some crucial factors like server location, energy grid used, training duration, and hardware specifications and uses them to estimate the amount of carbon emitted \cite{Lacoste2019QuantifyingCarbon}. Schmidt and others developed CodeCarbon, an open-source Python package that can be easily integrated into existing ML code to provide real-time estimates of carbon emissions during model training and inference \cite{Schmidt2021CodeCarbon}.

Since all these tools differ in methodology and produce different results, it has been not easy to compare the emissions reported by various researchers systematically. Moreover, there needs to be more consistency in what parts of a model's life cycle are considered while calculating these emissions. It is also important to note that the carbon emissions reported for most models only include the amount released to generate enough power for the hardware devices to perform model training and inference. The embodied emissions of ML models, like the ones due to hardware manufacturing, could be more well-known due to a lack of transparency and standard practices on the end of hardware manufacturers. Nevertheless, as we will see in this paper, despite the inconsistencies in calculations and reporting practices, we can still observe some strong trends in carbon emission levels across models.

\section{Research Questions}
The AI model training and tuning process comprises many different steps, each of which can influence the overall computational cost and environmental impact significantly. Thus, our goal in this work is to understand the predictors that impact or correlate with the carbon footprint of AI models. Even when not causal, identifying significant correlations helps uncover patterns and relationships that offer insights into the broader carbon footprint landscape. This understanding is crucial as it enables the identification of actionable steps that can be taken to reduce the negative environmental impact.

The factors we look in our work include the size and structure of the model (e.g., the number of parameters), the frequency of updates or commits, the repository’s age, and the organizational context (e.g., whether the model is developed by universities, companies, non-profits, or community-driven projects). Additionally, we investigate domain-specific differences across application areas such as Natural Language Processing (NLP), Computer Vision, and Audio. These factors motivate the following main research questions that we will tackle in this work.

\textbf{RQ1: What are the key predictors of carbon emissions in AI models?} This question seeks to identify the model- and repository-level attributes, such as the number of parameters, repository age, and development activity (e.g., commits), that significantly influence $CO_2$ emissions. As mentioned earlier, the goal of achieving "generality," i.e., creating models that are not task-specific and can perform any general task, has led to the race of training larger and larger models. Understanding how carbon emissions are related to the model sizes is important because it helps us understand the environmental impact of scaling the AI models and helps inform policy-making around large-scale AI developments. It also enables informed decisions on developing new model architectures, modifying previous architectures, and potential efficiency/accuracy trade-offs that should be taken. Similarly, repository age and commit activity provide valuable insights into the lifecycle and iterative development of models, reflecting their computational demands over time. Analyzing these factors can guide developers in adopting sustainable practices, while also identifying potential inefficiencies in long-term model maintenance and versioning.

\textbf{RQ2: How do different application domains, such as Natural Language Processing (NLP), Computer Vision, and Audio, impact the carbon footprint of AI models?}

This question aims to explore how varying application domains influence carbon emissions, recognizing that the computational demands differ significantly between tasks. For instance, intuitively, NLP models often require extensive pre-training but may exhibit lower emissions during fine-tuning compared to audio models, which frequently involve complex preprocessing and feature extraction. By examining these domain-specific trends, we can identify high-impact areas and guide the development of task-specific optimizations. Understanding these differences is essential for tailoring sustainability efforts to the unique needs and challenges of each domain. Studying the carbon trends against the task type is essential because it helps identify which applications have the highest environmental impact (and hence accurately forecast the impact across industries) and guides efforts toward optimizing high-emission task types. This also informs the development of specialized and energy-efficient hardware for specific tasks and uncovers domain-specific trends and challenges.

\textbf{RQ3: What role does organizational affiliation (e.g., universities, non-profits, companies, and community-driven projects) play in determining the carbon footprint of AI models?}
Organizational context has the potential to significantly shape the development and deployment practices of AI models, influencing their environmental impact. For example, university-driven projects often prioritize exploratory research, which may result in higher emissions due to experimental, large-scale training runs. Non-profits may focus on resource-intensive models aimed at public good, while companies might balance computational costs with profitability. Community-driven projects, on the other hand, often emphasize efficiency due to limited resources. By investigating these variations, we want to understand how different organizational practices would contribute to or mitigate carbon emissions. While understanding carbon footprints, it is important to understand how these systemic factors play a role as it can inform practices that each of these community groups should entail and these insights can also promote collaborative initiatives between different communities.

\textbf{RQ4: How can model popularity, as measured by likes and community engagement, serve as a proxy for environmental impact?}

This question explores whether popular models with high visibility and adoption rates tend to have a higher environmental cost. This can help better understand how user demand and community engagement influence computational resource utilization. No matter how large the models we train, at the end of the day, the environmental impact would be better gauged based on the usage of these models. Thus, it is crucial to study how the emissions vary by model popularity. This reveals the real-world environmental impact of widely-used AI models and enables prioritization of optimization efforts for models with the most extensive footprints. Moreover, it increases awareness about the hidden costs of the API-based AI services people use and encourages them to choose more sustainable AI solutions in the long run.

\textbf{RQ5: What best practices can be derived from this analysis to minimize the carbon footprint of AI models?}

The findings from any of the above research questions won't be useful if they cannot be translated into actionable outcomes and insights for the stakeholders in AI development and deployment. Thus, we bridge this gap by proposing specific recommendations and practices that stakeholders can adopt. For example, findings about the impact of model size and iterative development cycles can lead to policies encouraging more efficient training workflows and the use of parameter-efficient architectures. Insights about domain-specific emissions (e.g., NLP vs. audio models) could inform hardware optimizations tailored to different tasks. We discuss the exact findings and the proposed policy suggestions later in this paper.

\section{Data Analysis and Methods}
In this section, we share details about our data, its collection strategy and the method that we employed to study the trends.
\subsection{Dataset}
HuggingFace has become a widely used platform for researchers and developers to host their models. These models are accompanied by model cards that provide helpful information about the models. One such piece of information is equivalent carbon emissions, which were specially introduced to keep track of the environmental impact of the work in our field. The following is the proposed structure for the carbon emission metadata in the model cards:

\begin{Code}
co2_eq_emissions:
  emissions: number (in grams of CO2)
  source: "source of the information, either directly from AutoTrain, code carbon or from a scientific article documenting the model"
  training_type: "pre-training or fine-tuning"
  geographical_location: "as granular as possible, for instance Quebec, Canada or Brooklyn, NY, USA. To check your compute's electricity grid, you can check out https://app.electricitymap.org."
  hardware_used: "how much compute and what kind, e.g. 8 v100 GPUs"
\end{Code}

We scrape the hugging face model cards with an initial commit date from January 2019 to June 2024. This data was filtered based on the availability of required fields because many authors did not comply with the requirements for these model cards. The fields of our interest are as follows:

\begin{itemize}
\item Parameters: The number of trainable parameters in the model.
\item Likes: Popularity indicators (e.g., community engagement metrics like repository stars or likes).
\item Commits to Date: The total number of commits associated with the repository.
\item Repository Age: The duration (in years) since the repository was created.
\item Task Type: A categorical variable describing the type of machine learning task, such as "text-to-image" or "fill-mask".
\item Organizational Affiliation: A categorical variable capturing additional metadata about the repository’s focus, such as university or company.
\item Equivalent Carbon Emissions: Log-transformed estimates of the equivalent carbon emissions associated with training the model.
\end{itemize}

\begin{table*}[t]
\centering
\begin{tabular}{|p{4cm}|p{4cm}|p{4cm}|p{4cm}|}
\hline
\textbf{Task Type} & \textbf{Task Domain} & \textbf{Task Type} & \textbf{Task Domain} \\ \hline
'image-text-to-text' & Multimodal & 'summarization' & Natural Language Processing \\ \hline
'text-generation' & Natural Language Processing & 'visual-question-answering' & Multimodal \\ \hline
'audio-to-audio' & Audio & 'translation' & Natural Language Processing \\ \hline
'image-to-3d' & Computer Vision & 'sentence-similarity' & Natural Language Processing \\ \hline
'image-to-image' & Computer Vision & 'text-to-audio' & Audio \\ \hline
'fill-mask' & Natural Language Processing & 'text-to-video' & Computer Vision \\ \hline
'document-question-answering' & Multimodal & 'text-to-3d' & Computer Vision \\ \hline
'question-answering' & Natural Language Processing & 'robotics' & Reinforcement Learning \\ \hline
'image-to-video' & Computer Vision & 'text-to-speech' & Audio \\ \hline
'zero-shot-classification' & Natural Language Processing & 'text-to-image' & Computer Vision \\ \hline
'table-question-answering' & Natural Language Processing & 'audio-classification' & Audio \\ \hline
'text2text-generation' & Natural Language Processing & 'tabular-regression' & Tabular Data \\ \hline
'mask-generation' & Computer Vision & 'tabular-classification' & Tabular Data \\ \hline
'zero-shot-object-detection' & Computer Vision & 'time-series-forecasting' & Tabular Data \\ \hline
'image-feature-extraction' & Computer Vision & 'object-detection' & Computer Vision \\ \hline
'voice-activity-detection' & Audio & 'graph-ml' & Graph ML \\ \hline
'image-classification' & Computer Vision & 'automatic-speech-recognition' & Audio \\ \hline
'zero-shot-image-classification' & Computer Vision & 'video-classification' & Computer Vision \\ \hline
'image-to-text' & Computer Vision & 'token-classification' & Natural Language Processing \\ \hline
'unconditional-image-generation' & Computer Vision & 'text-classification' & Natural Language Processing \\ \hline
'image-segmentation' & Computer Vision & 'depth-estimation' & Computer Vision \\ \hline
'feature-extraction' & Natural Language Processing & 'reinforcement-learning' & Reinforcement Learning \\ \hline
\end{tabular}
\caption{Task Categories and Their Mappings}
\label{table:category-mapping}
\end{table*}

\subsection{Data Preprocessing}
Initial preprocessing steps included identifying which entries from the scraped data had the equivalent carbon emissions emitted reported. A custom function was implemented to clean and standardize string-based carbon emission values by removing suffixes like "g", "grams of CO2", etc. in the carbon emission values. To address the missing carbon emission values in the dataset, we implemented a multi-step approach to extract relevant information from the associated textual data, even for entries which had equivalent carbon emissions reported. We first used regex to split the text into multiple segments based on the presence of \#\# substring, which is the expected section divider as per the hugging face template. Then, we extract relevant sentences as those containing the word "environmental impact" and either "CO2" or "carbon" in a case-insensitive manner. If such a sentence was found, it was extracted as a potential source of carbon emission data. We further utilized a few-shot prompt-based approach using a GPT 4o large language model to identify carbon emission values explicitly stated in the extracted sentence. The prompt converted any emission values presented in kilograms to grams for consistency. If no carbon emission data was found in the text, the prompt instructed the model to return -1. We also provided a few examples to the model, including the cases when no relevant value was extracted in the original text, or the value was present in different units, or when the None output was returned by the regex function.

The exact prompt that was provided is as follows:

\textit{Given the data, what is the carbon emission in grams? If it is not present in the data, return -1. Remember that 1 kilogram = 1000 grams. Only output the final numerical value, no extra text. Data is here:}
\newline
\textbf{\{data\}}
\newline
\textit{Answer is here:}
\newline
\textbf{\{answer\}}
\newline
where data is the extracted sentence and answer (provided for the examples) is the carbon emission amount.

For entries where the carbon emission value was directly reported, we compared the value with the extracted values from this process, and found that there were discrepancies. Although the hugging face template suggests using the values in the column to be in grams, we saw many entries that did not abide by the suggestion as the value they reported in the text was the same numerical entry but in different units. In those cases, we overwrote the carbon emission value with the one provided in the text after converting it to proper units.

Additional features, such as parameters and likes, were converted to numeric types to ensure consistency. A mapping was applied to classify task types into broader task domains (e.g., Natural Language Processing, Computer Vision, etc.), and missing or unmatched task types were categorized as "Other." The mapping is shown in table \ref{table:category-mapping}. For organizational tags, a function extracted particular tags (e.g., non-profit, community, university, company), with unmatched tags labeled as "other." Finally, the repository's age was calculated in days by comparing the initial commit timestamp with the last commit timestamp.

\begin{figure*}[!ht]
  \centering
  \begin{minipage}[b]{0.49\textwidth}
    \centering
    \includegraphics[width=\linewidth]{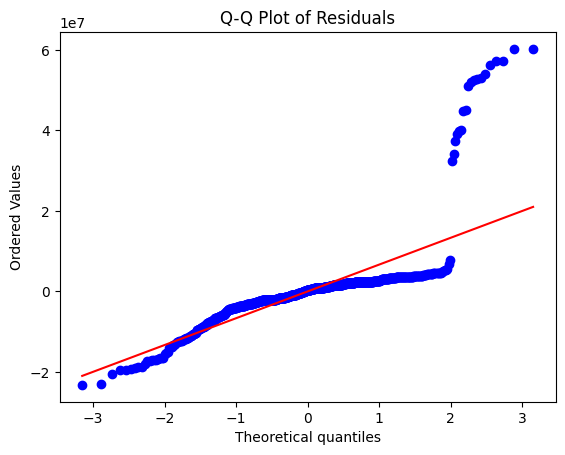}
    \caption{Q-Q plot of residuals before log transformation of the response variable.}
    \label{figure:qq-1}
  \end{minipage}
  \hfill
  \begin{minipage}[b]{0.49\textwidth}
    \centering
    \includegraphics[width=\linewidth]{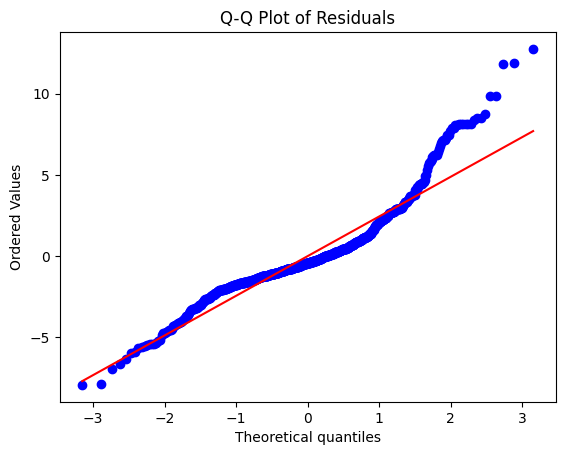}
    \caption{Q-Q plot of residuals after log transformation of the response variable.}
    \label{figure:qq-2}
  \end{minipage}
  \label{fig:qq-comparison}
\end{figure*}

\begin{figure*}[h]
  \centering
  \includegraphics[width=\linewidth]{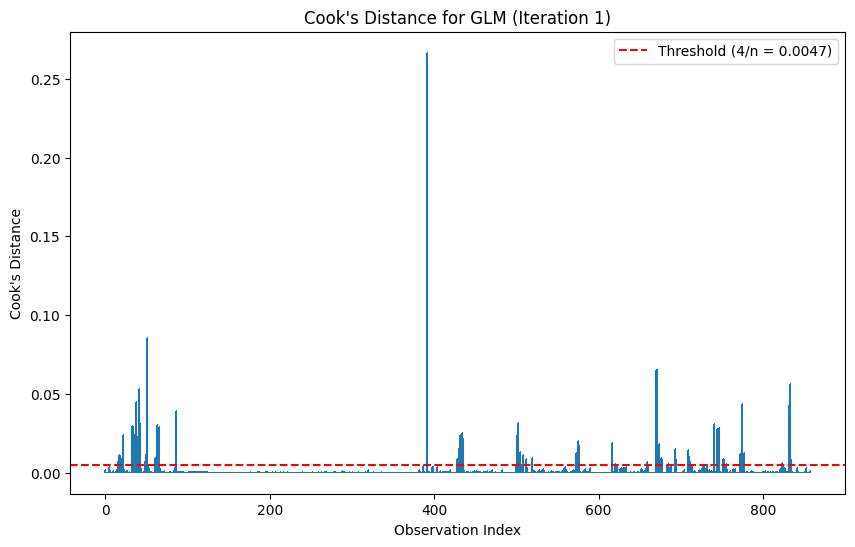}
  \caption{Cook's Distance plot for GLM (Iteration 1), highlighting influential observations exceeding the threshold (4/n = 0.0047) marked by the red dashed line.}
  \label{figure:cook-1}
\end{figure*}

\subsection{Assumptions}
Generalized Linear Models (GLMs) rely on several key assumptions for validity. For each assumption, we describe the tests conducted to evaluate them and the corresponding findings of our analysis.

\subsubsection{Independence of Observations}
Observations in the dataset must be independent of each other to ensure unbiased estimates and valid inference. For our dataset, independence is assumed due to the diverse sources of data, which includes contributions from multiple authors, spanning different time periods, and covering various task types across distinct datasets. This diversity inherently reduces the likelihood of dependency among observations and thus, the independence was assumed based on the nature of the data collection process.

\subsubsection{Appropriate Distribution of the Response Variable}
The assumption of an appropriate distribution for the response variable ensures that the data aligns with the specified distribution family, which in this case is Gaussian. Initially, the residuals of the response variable (equivalent carbon emissions) were examined using a Q-Q plot, which revealed deviations from normality, particularly at one tail-end as shown in figure \ref{figure:qq-1}. To address this, the response variable was log-transformed to better fit the Gaussian assumption, effectively reducing the impact of extreme values and improving distribution symmetry. A second Q-Q plot of the residuals post-transformation indicated substantial improvement with the residuals closely aligning with the theoretical normal line as shown in \ref{figure:qq-2}. This confirmed that the log transformation successfully adjusted the response variable to meet the distribution assumption, ensuring the validity of the GLM results.

\subsubsection{No Undue Influence from Outliers or High Leverage Points}
The assumption that outliers or influential points should not disproportionately affect the model is critical for ensuring reliable parameter estimates. Cook’s distance is a widely used diagnostic measure to identify such points. It quantifies the influence of each observation on the model's fitted values, combining information about the leverage of an observation and the residual error. We calculated Cook’s distance for all observations, and the observations with the distance exceeding the threshold 4/n (where n is the number of observations) were flagged as potentially influential. To address the issue, we iteratively removed these high-leverage points and refit the model at each step. This iterative approach ensures that the influence of any remaining points does not distort the final model estimates because at each iteration, we recheck Cook’s distance to confirm whether additional influential points persisted. Figure \ref{figure:cook-1} shows the cook's distance plot from the very first iteration. The 17th iteration resulted in a dataset with no outlier/extreme points.

\subsubsection{No Multicollinearity}

The assumption of no multicollinearity ensures that predictors are not highly correlated, as this can inflate the variance of estimates and reduce model interpretability. To assess multicollinearity, the Variance Inflation Factor (VIF) was calculated for each predictor. A threshold of VIF > 5 was used to flag predictors with high collinearity for removal.

Initial results revealed that predictors such as “Computer Vision” (VIF = 7.69) and “Natural Language Processing” (VIF = 7.63) exhibited multicollinearity issues. Given that the number of data points for computer vision (95) were considerably less compared to the number of data points for natural language processing (741), we removed the computer vision predictor from the model to improve stability and interpretability. After the removal, the VIF values for all remaining predictors dropped below the threshold, with the highest being the intercept (const) at 331.49, which is expected and does not impact the interpretation of other predictors.

\subsection{Data Modeling}
The Generalized Linear Model (GLM) is a flexible extension of ordinary linear regression, enabling analysis of relationships between a dependent variable and one or more independent variables under a broader range of distributions. GLMs are composed of three main components:

\begin{itemize}
    \item Random Component: Specifies the probability distribution of the dependent variable. For this study, the dependent variable—log-transformed equivalent carbon emissions—is assumed to follow a Gaussian (normal) distribution.
    \item Systematic Component: Represents the linear predictor, a linear combination of the independent variables and their corresponding coefficients.
    \item Link Function: Connects the expected value of the dependent variable to the linear predictor. In this case, an identity link function is used, implying that the expected value of equivalent carbon emissions directly corresponds to the linear predictor.
\end{itemize}

In this work, the pipeline category and organization tag variables were encoded using one-hot encoding to create binary indicator variables for each unique category. Categories with low representation or other tag were excluded to avoid potential noise and multicollinearity. Continuous variables (parameters, likes, commits to date) were log-transformed using base-10 logarithms to reduce skewness and stabilize variance. A small constant (0.001) was added to avoid undefined values due to logarithms of zero. The repository age variable was retained in its original scale. All transformed and encoded features were combined into a single design matrix and a constant term was added to the model matrix to account for the intercept term in the GLM.

\begin{table*}[t!]
\caption{Generalized Linear Model Regression Results}
\centering
\label{table:glm-results}
\begin{tabular}{@{}ll@{}}
\toprule
\textbf{Model:}               & GLM                                \\
\textbf{Dependent Variable:}  & $y$                                \\
\textbf{No. Observations:}    & 782                                \\
\textbf{Method:}              & IRLS                               \\
\textbf{Model Family:}        & Gaussian                           \\
\textbf{Df Residuals:}        & 771                                \\
\textbf{Df Model:}            & 10                                 \\
\textbf{Link Function:}       & Identity                           \\
\textbf{Scale:}               & 2.2778                             \\
\textbf{Log-Likelihood:}      & -1425.9                            \\
\textbf{Deviance:}            & 1756.2                             \\
\textbf{Pseudo R-squ. (CS):}  & 0.7732                             \\
\textbf{Pearson chi2:}        & 1.76e+03                           \\
\textbf{No. Iterations:}      & 3                                  \\
\textbf{Covariance Type:}     & nonrobust                          \\
\midrule
\end{tabular}

\vspace{1em}

\begin{tabular}{@{}lrrrrr@{}}
\toprule
\textbf{}         & \textbf{Coef.} & \textbf{Std.Err.} & \textbf{z} & \textbf{P>|z|} & \textbf{[0.025, 0.975]} \\
\midrule
const             & -6.0102       & 0.983             & -6.117      & 0.000          & [-7.936, -4.084]        \\
parameters        & 0.8740        & 0.113             & 7.725       & 0.000          & [0.652, 1.096]          \\
likes             & 0.1309        & 0.056             & 2.357       & 0.018          & [0.022, 0.240]          \\
commits\_to\_date  & 2.2305        & 0.088             & 25.208      & 0.000          & [2.057, 2.404]          \\
repo\_age         & 0.0022        & 0.001             & 4.126       & 0.000          & [0.001, 0.003]          \\
Audio             & 7.9926        & 0.769             & 10.400      & 0.000          & [6.486, 9.499]          \\
Natural Language Processing & -0.3789 & 0.179             & -2.112      & 0.035          & [-0.731, -0.027]        \\
community         & -2.2625       & 0.628             & -3.600      & 0.000          & [-3.494, -1.031]        \\
company           & 0.4205        & 0.263             & 1.602       & 0.109          & [-0.094, 0.935]         \\
non-profit        & 1.1921        & 0.330             & 3.615       & 0.000          & [0.546, 1.838]          \\
university        & 4.6197        & 0.506             & 9.126       & 0.000          & [3.628, 5.612]          \\
\bottomrule
\end{tabular}
\end{table*}

\section{Results and Discussion}
The results of the Generalized Linear Model (GLM) analysis, as shown in table \ref{table:glm-results}, indicate several key predictors of carbon equivalent emissions (log-transformed) for AI models. The model demonstrates a pseudo R-squared (Cragg \& Uhler’s) value of 0.7732, indicating a substantial proportion of the variance in equivalent carbon emissions is explained by the predictors. Below is a detailed discussion of the findings categorized by core predictors, pipeline categories, and organizational affiliation. The model’s low deviance (1756.2) and rapid convergence in three iterations indicate robust performance and stable parameter estimation. These metrics validate the reliability of the findings and underscore the significance of the predictors in explaining equivalent carbon emissions. We use a threshold of 0.05 for the $p$-value.

\subsection{Core Predictors}
\subsubsection{Model Parameters}
The number of parameters in the model shows a positive association with carbon emissions ($\beta$ = 0.874, $p$ < 0.001). Larger models require increased computational resources, resulting in higher emissions. These findings highlight the need to prioritize parameter-efficient techniques such as pruning, knowledge distillation, and low-rank approximations to mitigate environmental impact. Furthermore, the trend suggests that the race toward larger, more generalized models could exacerbate carbon emissions if sustainable practices are not adopted. Developers and researchers need to carefully balance model performance with environmental costs by exploring techniques that achieve comparable outcomes with fewer parameters.

\subsubsection{Commits to Date}
The number of commits exhibits a strong positive relationship with emissions ($\beta$ = 2.2305, $p$ < 0.001). Frequent updates and iterative development cycles require substantial computational resources, as each cycle often involves retraining or fine-tuning models. While this reflects active development and improvement, it also points to inefficiencies that could be addressed by adopting energy-efficient workflows, optimizing retraining strategies, and limiting unnecessary updates. Encouraging practices like batch updates or modular design could reduce the frequency of resource-intensive updates while maintaining model quality.

\subsubsection{Model Popularity}
The number of likes positively correlates with emissions ($\beta$ = 0.1309, $p$ = 0.018), indicating that more popular models may be extensively retrained or fine-tuned by the community, leading to higher emissions. This highlights the need for responsible usage of popular models, particularly by organizations and developers who widely adopt these models. By optimizing popular models for energy efficiency, stakeholders can significantly reduce their real-world carbon impact. Additionally, this finding raises awareness about the hidden environmental costs of community-driven engagement with AI systems.

\subsubsection{Repository Age}
Repository age shows a small but significant positive association with emissions ($\beta$ = 0.0022, $p$ < 0.001). Older repositories may involve long-term updates and maintenance, which could contribute to emissions. This suggests that while older repositories are often associated with more mature and optimized workflows, prolonged maintenance cycles can still add to their environmental footprint. Encouraging efficient archival practices and reducing the frequency of unnecessary updates for older repositories could help mitigate this impact.

\subsection{Task Domains}
\subsubsection{Audio Models}
Audio models display the highest positive effect on emissions ($\beta$ = 7.9926, $p$ < 0.001). This significant result might reflect the computational demands of processing time-series data in audio tasks or may simply be a result of the unavailability of large-scale data entries in the dataset. Audio tasks often require resource-intensive preprocessing and feature extraction, which add to their carbon footprint. Task-specific optimizations, such as efficient encoding and processing pipelines, and exploring lightweight architectures could significantly reduce emissions in this domain. The findings suggest prioritizing sustainability efforts in audio-based AI development.

\subsubsection{Natural Language Processing (NLP) Models}
NLP models exhibit a small but significant reduction in emissions ($\beta$ = -0.3789, $p$ = 0.035). This result likely reflects the efficiency of fine-tuning pre-trained transformers rather than training models from scratch. Fine-tuning enables developers to leverage the computational effort already invested in pre-training large models, resulting in lower overall emissions. This finding underscores the importance of reusing pre-trained architectures and adopting task-specific fine-tuning as a sustainable practice in NLP development.

\subsection{Organizational Affiliation}
\subsubsection{University}
University-affiliated repositories show the highest emissions ($\beta$ = 4.6197, $p$ < 0.001). This likely reflects the exploratory nature of academic research, which often involves resource-intensive experimental models. Universities could adopt sustainability practices such as using renewable energy sources in training pipelines and incorporating green AI principles into their research methodologies. Encouraging collaboration among academic institutions to share computational resources could also help minimize emissions without hindering innovation.

\subsubsection{Non-Profit}
Non-profit organizations are associated with significantly higher emissions ($\beta$ = 1.1921, $p$ < 0.001). These organizations often train large models for societal benefits, such as public good or open-access research. To balance their objectives with environmental sustainability, non-profits should focus on energy-efficient model training and explore shared resources or collaborations with other sectors to optimize computational demands.

\subsubsection{Company}
Company-affiliated repositories show a moderate but insignificant positive association with emissions ($\beta$ = 0.4205, $p$ = 0.109). Companies often prioritize balancing computational costs with profitability, which could explain the moderate emissions observed. This provides an opportunity for companies to lead sustainability efforts by adopting eco-friendly AI practices, such as investing in energy-efficient hardware, utilizing renewable energy, and creating standardized guidelines for sustainable model development.

\subsubsection{Community}
Community-driven repositories exhibit significantly lower emissions ($\beta$ = -2.2625, $p$ < 0.001). This result highlights the efficiency-focused development practices often adopted in community projects due to resource constraints. These projects demonstrate that impactful AI models can be developed with limited resources, offering a blueprint for reducing emissions across other sectors. Supporting community-driven initiatives and promoting collaboration between communities and organizations could amplify these benefits and drive further innovation in sustainable AI practices.

\section{Limitations}
While this work provides valuable initial insights into the factors influencing the carbon footprint of AI models, several limitations must be acknowledged. First, the findings are based on a specific dataset and may not generalize to all AI development contexts or applications. The dataset’s coverage of certain domains, such as audio models, may be limited, potentially biasing the results. Additionally, the analysis relies on observational data, which establishes correlations rather than causal relationships. For instance, while predictors like model popularity or organizational affiliation are associated with emissions, the exact mechanisms driving these relationships require further exploration. The scope of this analysis is primarily focused on quantifiable attributes (e.g., parameters, commits), and qualitative factors like organizational priorities or developer behavior remain unexplored.
\section{Next Steps}
We believe that this work lays a fundamental groundwork for understanding the environmental impact of AI models. It establishes a framework for identifying key predictors of emissions and highlights the need for more sustainable AI practices. Future work should build upon these initial findings by incorporating larger and more diverse datasets, enabling a broader generalization across different domains, organizations, and geographies.

Future research should also aim to uncover causal mechanisms underlying the observed correlations, leveraging advanced methodologies and experimental designs. This would provide stronger evidence for proposed intuitive reasonings, targeted interventions and policy recommendations. Additionally, exploring qualitative factors, such as organizational decision-making processes and societal influences, could complement the quantitative analysis and provide a holistic perspective on AI sustainability.

Long-term efforts should focus on developing standardized metrics for emissions, accounting for regional energy differences and hardware efficiency variations. Collaborative initiatives involving academia, industry, and the community will be critical for scaling sustainable practices and fostering collective accountability. By addressing these challenges, we hope to inspire a comprehensive research agenda that aligns AI innovation with ecological responsibility.

\section{Conclusion}
This work provides an initial exploration into the factors influencing the carbon footprint of AI models, identifying key predictors such as model parameters, commits, and organizational contexts, as well as domain-specific trends. By establishing a framework for analyzing emissions, this study underscores the environmental implications of AI development and highlights actionable areas for improvement, such as adopting parameter-efficient architectures and optimizing development workflows. While the findings are preliminary, they lay the foundation for future research to build upon, enabling deeper insights and more effective strategies for sustainable AI practices.

\bibliographystyle{ACM-Reference-Format}
\bibliography{main}










\end{document}